\documentclass[times,twocolumn,final]{elsarticle}

\usepackage{medima}
\usepackage{framed,multirow}

\usepackage{amssymb}
\usepackage{latexsym}

\usepackage{url}
\usepackage{xcolor}
\usepackage{enumerate}

\usepackage{enumitem}
\usepackage{amsmath}
\usepackage{booktabs}
\usepackage{multirow}
\usepackage{hyperref}

\definecolor{newcolor}{rgb}{.8,.349,.1}

\journal{Medical Image Analysis}

\begin{document}

\verso{Senmao Wang \textit{et~al.}}

\begin{frontmatter}

\title{Costal Cartilage Segmentation with Topology Guided Deformable Mamba: Method and Benchmark}

\author[1]{Senmao \snm{Wang}\fnref{fn1}}
\author[2]{Haifan \snm{Gong}\fnref{fn1}}
\author[1]{Runmeng \snm{Cui}\fnref{fn1}}
\author[3]{Boyao \snm{Wan}}
\author[1]{Yicheng \snm{Liu}}
\author[1]{Zhonglin \snm{Hu}}
\author[4]{Haiqing \snm{Yang}}
\author[1]{Jingyang \snm{Zhou}}
\author[1]{Bo \snm{Pan}}
\author[1]{Lin \snm{Lin}\corref{cor2}}
\author[1]{Haiyue \snm{Jiang}\corref{cor2}}
\fntext[fn1]{Joint first authors}
\cortext[cor2]{Corresponding authors}
\ead{linlin@psh.pumc.edu.cn (Lin Lin), haiyuejiang@psh.pumc.edu.cn}
\address[1]{Plastic Surgery Hospital, Chinese Academy of Medical Sciences and Peking Union Medical College, Beijing, 100144, China} 
\address[2]{School of Science and Engineering, The Chinese University of Hong Kong, Shenzhen, 518172, China}
\address[3]{Nanjing University of Science and Technology, Nanjing, 210094, China}
\address[4]{The Second Hospital of Hebei Medical University, Shijiazhuang, 050000, China}

\received{1 May 2013}
\finalform{10 May 2013}
\accepted{13 May 2013}
\availableonline{15 May 2013}
\communicated{S. Sarkar}

\begin{abstract}
Costal cartilage segmentation is crucial to various medical applications, necessitating precise and reliable techniques due to its complex anatomy and the importance of accurate diagnosis and surgical planning. We propose a novel deep learning-based approach called topology-guided deformable Mamba (TGDM) for costal cartilage segmentation. The TGDM is tailored to capture the intricate long-range costal cartilage relationships. Our method leverages a deformable model that integrates topological priors to enhance the adaptability and accuracy of the segmentation process. Furthermore, we developed a comprehensive benchmark that contains 165 cases for costal cartilage segmentation. This benchmark sets a new standard for evaluating costal cartilage segmentation techniques and provides a valuable resource for future research. Extensive experiments conducted on both in-domain benchmarks and out-of-domain test sets demonstrate the superiority of our approach over existing methods, showing significant improvements in segmentation precision and robustness. All code and data will be made publicly available after acceptance.

\end{abstract}


\end{frontmatter}


\section{Introduction}
\label{sec:introduction}

The costal cartilage (CC) is a crucial component of the thoracic wall structure. Specifically, the first to tenth CCs connect the ribs to the sternum, while the seventh to tenth CCs contribute to forming the costal arch. The eleventh and twelfth ribs terminate in the abdominal muscles and do not attach to the sternum, with their tips possibly lacking cartilage. With advancing age, CC undergoes pathophysiological changes such as increased matrix calcification \citep{Lau2015} and fibrous formation \citep{Hukins1976}, resulting in increased computed tomography (CT) attenuation. The labeling and segmenting CC are essential for studying its anatomy and pathological changes and optimizing clinical application strategies (Figure 1). In addition, as a type of hyaline cartilage, CC is favored in plastic and reconstructive surgery due to its abundant availability and excellent mechanical properties, making it one of the most important autologous materials. \citep{Fisher2022}. Millions of people worldwide suffer from microtia—a congenital deformity that significantly impacts facial appearance and psychological well-being \citep{Zhang2018}. The gold standard treatment for microtia is ear reconstruction surgery using CC \citep{Wang2023}. This procedure typically involves harvesting portions of the sixth to ninth CCs to create the ear framework, but it carries risks of thoracic deformity and pain \citep{Wilkes2014}. Precise labeling and segmenting CC can facilitate personalized harvesting plans, potentially minimizing donor site damage and reducing complication risks \citep{Wang2024}.
\begin{figure}[!thbp]
\includegraphics[width=\linewidth]{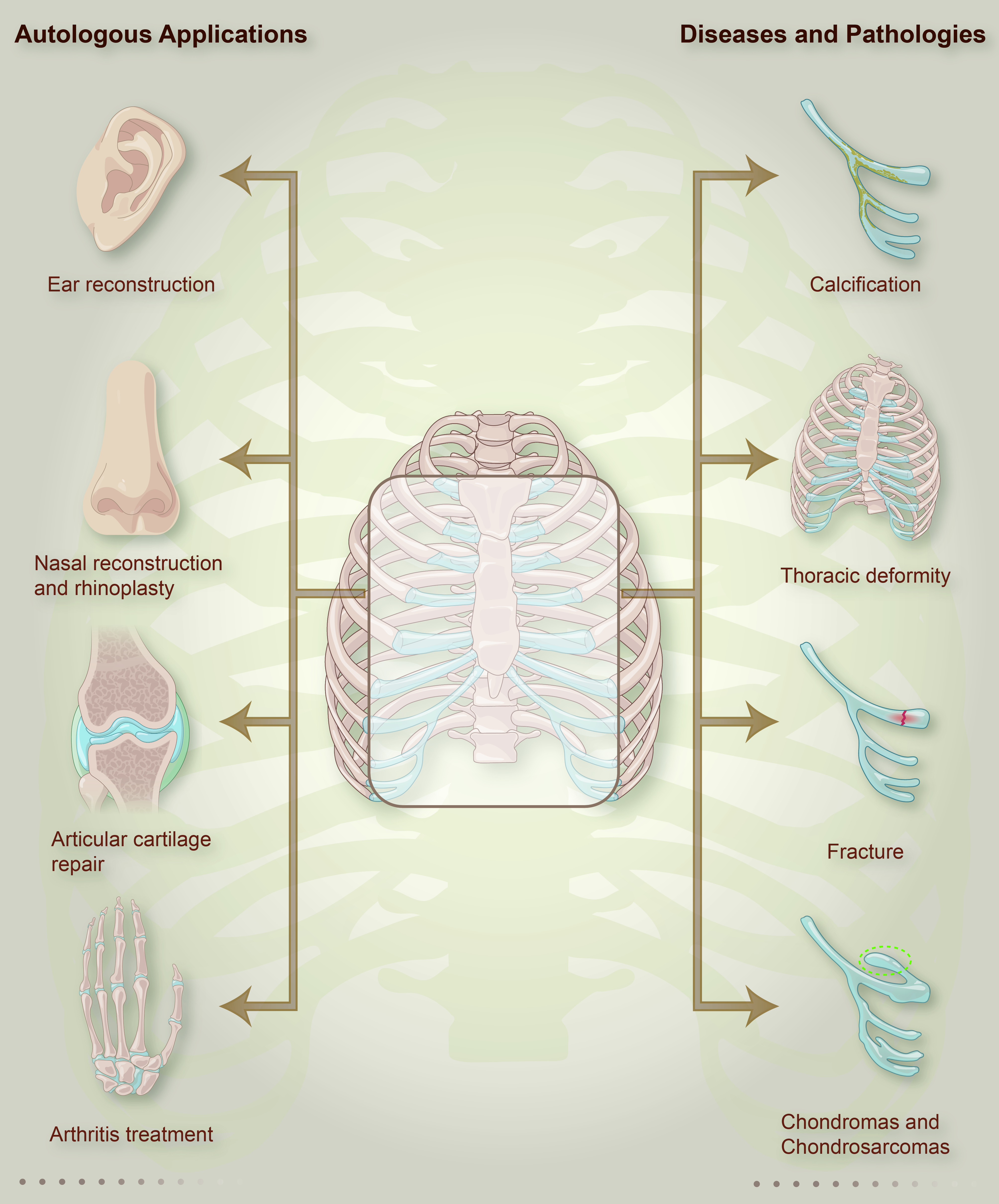}
\caption{Autologous applications and pathological changes of CC.} 
\label{datavis}
\end{figure}
Additionally, CC is pivotal for nasal reconstruction \citep{Chen2022}, orbital reconstruction \citep{Kajikawa2010}, articular cartilage repair \citep{Yoon2024}, arthritis treatment \citep{Pang2023}, and pectus excavatum repair \citep{Zhang2023}. Labeling and segmenting CC and creating personalized cartilage models for preoperative simulation are crucial for optimizing surgical outcomes. \citep{Wang2024}.

The segmentation and annotation of CC are also important for thoracic anatomy research and disease diagnosis and treatment. Various studies focus on lung volume \citep{Mansoor2014}, spinal \citep{Busch2023} and thoracic deformities \citep{Schlager2022}, and rib anatomy \citep{yang2021ribseg,Jin2023ribsegv2,yang2024deep}. However, \citet{Nummela2018} emphasized that machine learning-based research on the segmentation of CC, a crucial component of the thoracic wall, remains largely unexplored. Accurate segmentation of CC is essential for the precise diagnosis and classification of CC fractures, as noted by \citet{Nummela2018}, the diagnosis and delineation of resection boundaries for chondromas \citep{Liu2023} and chondrosarcomas \citep{Marco2000}, and providing valuable references for thoracic research \citep{Busch2023,Oliveira2018}. 

The advent of computer vision and deep learning has propelled automation to the forefront of medical imaging segmentation~\citep{isensee2021nnu,li2023lvit,shaker2024unetr++}. However, CC segmentation remains a challenging task due to the similar intensity of the foreground and background (e.g., organs), particularly in younger teenagers whose CC is  softer than those of adults~\citep{weber2021,gong2024intensity,Wang2024}. Additionally, the local receptive field of CNN-based methods~\citep{ronneberger2015u,isensee2021nnu} limits the ability of neural networks to segment CC accurately, which occupies a long and thin regions. To address these issues, we consider it crucial to embed long-range relationship modeling for the CC structure while accounting for the continuity of adjacent foreground areas. Mamba~\citep{gu2023mamba,mamba2} is utilized in our study due to its capability to construct long-distance dependencies with linear time complexity. This feature is particularly advantageous in modeling the intricate long-range dependencies among the voxels within the costal cartilage, which is crucial for the segmentation tasks in our deep learning model. Importantly, Mamba achieves this with linear time complexity, which ensures the efficiency of our model. Based on the above concerns, we propose a novel topology-guided deformable Mamba (TGDM) framework for CC segmentation. The backbone used here is a UNet-like structure with a two-stage segmentation process. In the first stage, the bounding box of the CC is segmented to obtain its coarse position, enhancing the reliability of the topology prior and modeling each CC's long-range relationship based on the relative coordinates of the foreground centerline. In the second stage, two critical components are introduced: position shifting mamba (PSM) and grouped deformable Mamba (GDM). PSM obtains the centerline coordinates for each costal cartilage, while GDM models the long-range relationships guided by the position of each CC. The contributions of this work are summarized as follows:
\begin{itemize}
  \item We introduce a novel segmentation approach termed TGDM, which leverages topological knowledge to enhance the segmentation of CC. This model is specifically tailored to understanding and modeling the complex long-range spatial relationships inherent in CC structures.
  \item We design two novel modules: PSM and GDM. PSM obtains topological prior knowledge, while GDM uses this prior knowledge to enhance costal cartilage segmentation by modeling the long-range deformable relationship on the foreground CC. These modules work together to model the complex long-range spatial relationships inherent in CC structures effectively.
  \item We develop and validate a comprehensive benchmark consisting of a large dataset of cases, which is used to rigorously test the effectiveness and robustness of our proposed method. This benchmark sets a new standard for future studies and method comparisons in the domain of CC segmentation.
  \item Our method demonstrates superior DSC and NSD scores on both in-domain and out-of-domain data, highlighting its robustness and adaptability. This is particularly crucial for medical applications where variability across patients and imaging conditions can significantly impact segmentation algorithm DSC and NSD scores.
\end{itemize}

\section{Related Works}

\subsection{Costal Cartilage Segmentation}
Segmentation of CC has gained attention due to its relevance in diagnostic and therapeutic applications in thoracic imaging. Early works, such as \citet{holbrook2007segmentation}, developed methods to segment CCs in abdominal CT using watershed markers, demonstrating the feasibility of image processing techniques for enhancing cartilage visibility. \citet{noorda2012segmentation} expanded upon this by segmenting the cartilage in the rib cage using 3D MRI, which highlighted differences in segmentation challenges across imaging modalities. \citet{barbosa2015semi} introduced a semiautomatic 3D segmentation method for CT data from patients with pectus excavatum that was tailored to specific clinical conditions that alter the thoracic structure. More recently, \citet{noguchi2020bone} applied convolutional neural networks (CNNs) with novel data augmentation techniques for bone segmentation on whole-body CT, indirectly contributing to the field by improving the segmentation reliability of bony structures adjacent to CC. These advancements reflect a broader trend toward automation and precision in segmenting thoracic anatomical structures. However, most of the aforementioned work is based on semiautomatic or traditional segmentation methods. Additionally, the lack of a large-scale dataset in this field has limited the development of deep learning-based segmentation methods. To address these issues, we compile a large-scale dataset in this paper.

\subsection{Medical Volume Segmentation}
The field of medical volume segmentation has undergone significant advancements by integrating deep learning technologies. Pioneering contributions by \citet{ronneberger2015u} with UNet and its later extension, nnU-Net by \citet{isensee2021nnu}, have set foundational standards for CNN-based approaches in handling complex biomedical image data. The application of these networks has been demonstrated in diverse challenges such as the multimodal brain tumor segmentation in \citet{pei2021multimodal}. \citet{roy2023mednext} proposed MedNext, which uses a large kernel to construct a long-range relationship. The emergence of transformer-based models has led to the introduction of novel methodologies that capitalize on the ability of these models to manage long-range dependencies, as seen in SwinUNETR \citep{hatamizadeh2021swin} and further developments in UNETR \citep{hatamizadeh2022unetr} , UNETR++ \citep{shaker2024unetr++}, and LViT \citep{li2023lvit}. nnFormer \citep{zhou2023nnformer} highlighted the ongoing shift towards integrating convolutional and transformer networks to enhance segmentation DSC scores. Additionally, the introduction of state space models~\citep{gu2023mamba} in segmentation tasks through U-Mamba and nnMamba by \citet{ma2024u} and \citet{gong2024nnmamba}, respectively, underscores the continuous evolution towards more efficient and accurate segmentation frameworks. These advancements collectively push the boundaries of what is possible in medical volume segmentation, promising further improvements in diagnostic, therapeutic, and surgical applications. Nevertheless, there are no segmentation methods that are tailor-designed for CC segmentation from CT images, which is a challenging task due to the similar contrast between the cartilage and other organs.  address this issue by using the low-resource ability of Mamba \citep{gu2023mamba,mamba2} to model the long range relationship.

\subsection{Prior Guided Segmentation}
Incorporating prior knowledge into deep learning models has significantly boosted the accuracy and robustness of medical image segmentation. \citet{huang2021medical} introduced deep atlas priors to enhance medical image segmentation by integrating anatomical priors into the learning process. Similarly, \citet{guo2022cardiac} demonstrated the effectiveness of combining deep learning uncertainty with shape priors for sparse annotation in cardiac MRI segmentation. \citet{gong2021multi,gong2023thyroid} utilized thyroid region priors to guide attention mechanisms for more accurate segmentation of thyroid nodule ultrasound. Additionally, \citet{pei2023pets} developed PETS-Nets, which are anatomy-guided networks for joint pose estimation and tissue segmentation, showcasing the integration of complex anatomical knowledge. The ADNet++ framework by \citet{hansen2023adnet++} employs a few-shot learning approach with uncertainty-guided feature refinement for multiclass medical image volume segmentation. \citet{zhao2023masked} investigate the frequency prior to guide the domain adaptive laparoscopic images segmentation task. Furthermore, \citet{lin2024dynamic} proposed dynamic-guided spatiotemporal attention for echocardiography video segmentation, emphasizing the value of dynamic information in temporal data. These advancements highlight the ongoing trend of using various forms of prior knowledge to refine segmentation processes, leading to more precise and clinically relevant outcomes. In this work, we introduce a novel deformable attention module that leverages the topological priors of CC to improve the long-range modeling capabilities of CNNs for foreground CC segmentation to enhance network precision and efficiency by integrating spatial context more effectively.

\begin{figure*}[thbp]
\centering
\includegraphics[width=0.96\linewidth]{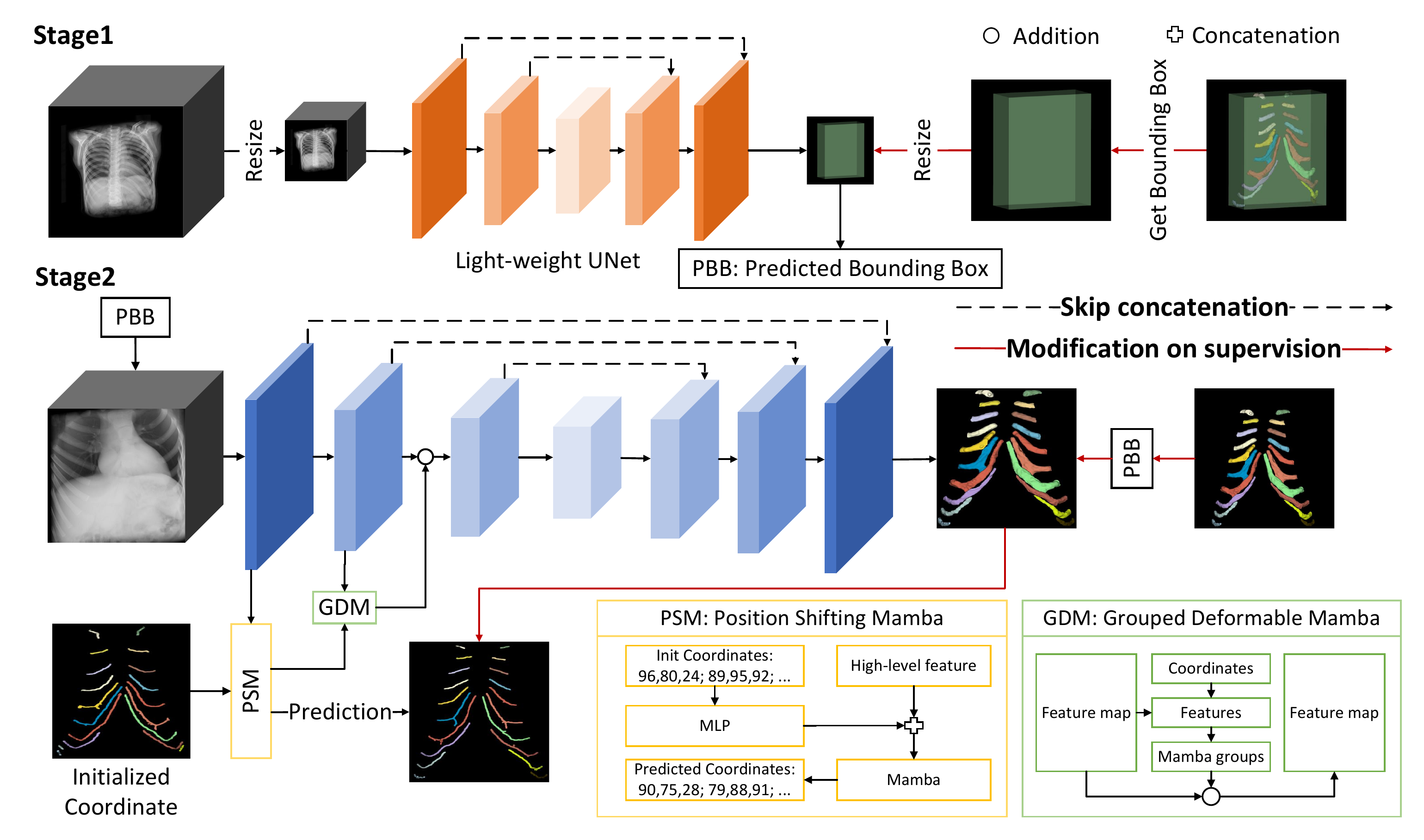}
\caption{Overall pipeline of the proposed method. The stage 1 extracts the bounding box of the CC to help the network identify the approximate region of the CC. The stage 2 involves semantic segmentation of the CC based on topological structure prior knowledge. Using the results from the first stage and the corresponding CC location templates, we obtain the coordinates of each CC centerline through a mamba. Then, based on these coordinates, we use mamba for feature extraction to achieve more efficient interaction results of CC features.}
\label{fig:pipeline}
\end{figure*}

\section{Methods}

\subsection{Overview}
The overall pipeline of the proposed method consists of two stages, as shown in Figure~\ref{fig:pipeline}.  In the first stage, the bounding box of the CC was extracted to help the network identify the coarse region of the CC. The second stage involves semantic segmentation of the CC based on topological structure prior knowledge. Using the results from the first stage and the corresponding CC location templates, we obtained the coordinates of each CC centerline through a Mamba. Then, based on these coordinates, we used Mamba for feature extraction to achieve more efficient results for the interaction of CC features.

\subsection{Coarse Segmentation with Light-weight U-Net}
In the first stage, we use a light-weight U-Net to obtain a coarse estimation of the position of the CC at a low resolution. This preliminary phase has two primary purposes: (1) to enhance the reliability of the initial coordinates by providing a rough yet consistent estimate of the cartilage location, and (2) to enable subsequent segmentation at a higher resolution, thereby refining the segmentation boundaries and improving the overall accuracy of the results.

Formally, let $\mathbf{X}_{\text{high}}$ represent the high-resolution input image and $\mathbf{Y}_{\text{high}}$ denote the ground truth (GT) segmentation at the original resolution. Based on the GT at the original resolution, we can obtain the GT bounding box. Using this bounding box, we construct a binarized segmentation pseudo-label to represent the approximate CC GT region. Since this region is solid, we can directly resize the bounding box of GT without the risk of losing the fine CC details. For this purpose, let $\mathbf{X}_{\text{low}}$ and $\mathbf{Y}_{\text{box}}$ represent the resized low-resolution image and the bounding box segmentation result, respectively. We use a lightweight U-Net, denoted as $\mathcal{F}_{\text{lightunet}}$, to segment this bounding box to obtain the approximate size of the CC. The initial low-resolution segmentation can be represented as:
\begin{equation}
\mathbf{S}_{\text{low}} = \mathcal{F}_{\text{lightunet}}(\mathbf{X}_{\text{low}}).
\end{equation}
To optimize this lightweight U-Net, we use a combination of Dice and Cross-Entropy (CE) losses. The total loss for optimizing the lightweight U-Net is:
\begin{equation}
\mathcal{L}_{\text{stage1}} =\mathcal{L}_{\text{Dice}}(\mathbf{S}_{\text{low}}, \mathbf{Y}_{\text{box}}) +\mathcal{L}_{\text{CE}}(\mathbf{S}_{\text{low}}, \mathbf{Y}_{\text{box}}).
\end{equation}
By employing a low-resolution training phase, we ensure that the initialized coordinates are more robust, which is essential for the subsequent high-resolution segmentation. This multi-resolution approach not only improves the initial localization but also is time-efficient comparied with the method that directly locates the binary CC. In stage 2, all the parameters in stage 1 are fixed, and the predicted bounding boxs are sent to the stage 2 for further precise CC segmentation.

\begin{figure*}[thbp]
\centering
\includegraphics[width=1.02\linewidth]{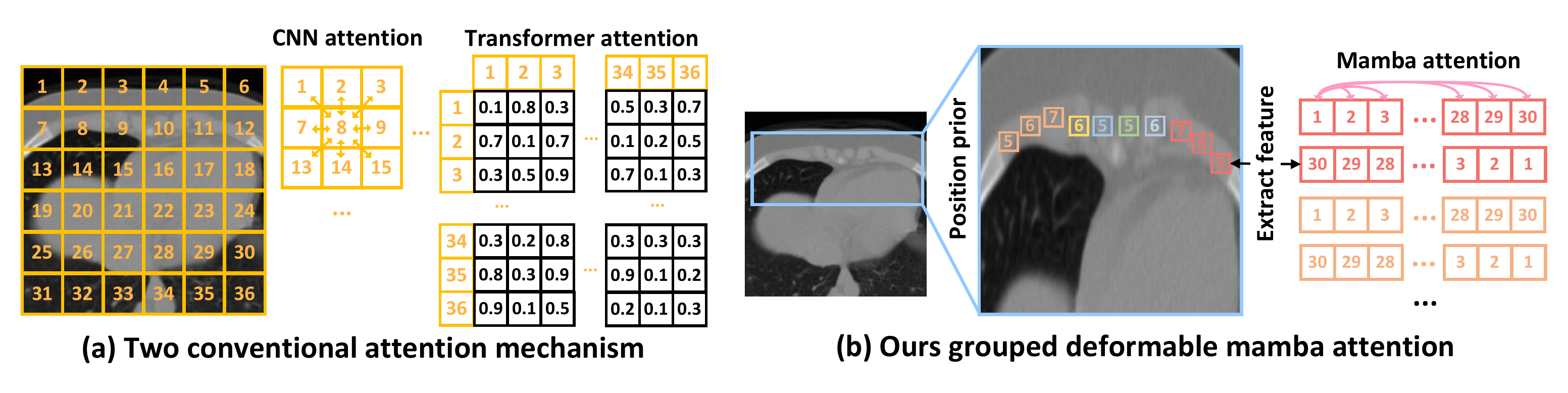}
\caption{The proposed grouped deformable mamba attention mechanism compared to previous sequential or self-attention mechanisms. Our deformable attention mechanism consists of two parts: coordinate acquisition based on position-prior and the grouped mamba attention mechanism.}
\label{fig:attn}
\end{figure*}

\subsection{Topology-Guided Deformable Mamba}

\textbf{Motivation:} Incorporating the deformable Mamba module in our framework is driven by the need to establish long-range dependencies within the CC structure, using its topology prior (i.e., centerline) as a guide. Segmenting the CC presents significant challenges due to the low contrast between the CC and the surrounding background tissues. Additionally, the CC typically maintains a relatively fixed anatomical position, which complicates the segmentation process. The deformable Mamba module addresses these challenges by effectively capturing and modeling the spatial relationships and variations along the CC, enhancing  segmentation accuracy and robustness. This module leverages the centerline to provide a structural context, allowing for more precise identification and delineation of the CC against the low-contrast background.

\subsubsection{Mamba: A Powerful State Space Model at Linear Complexity:} Before detailing our deformable Mamba module, we first introduce Mamba~\citep{gu2023mamba,mamba2}, an advanced state space model designed for long-range relationship modeling.  The selective state space model (SSM) in Mamba needs four input-dependent parameters: $(\Delta, \mathbf{A}, \mathbf{B}, \mathbf{C})$.  Then Mamba transforms them into $(\overline{\mathbf{A}}, \overline{\mathbf{B}}, \mathbf{C})$ using the following equations:
\begin{equation}
\overline{\mathbf{A}} = \exp(\Delta \mathbf{A}), \quad \overline{\mathbf{B}} = (\Delta \mathbf{A})^{-1} (\exp(\Delta \mathbf{A}) - \mathbf{I}) \cdot \Delta \mathbf{B}.
\end{equation}
The sequence-to-sequence transformation of the SSM can then be expressed as:
\begin{equation}
\begin{aligned}
h_t &= \overline{\mathbf{A}} h_{t-1} + \overline{\mathbf{B}} x_t, \\
y_t &= \mathbf{C} h_t,
\end{aligned}
\end{equation}
where \( t \) denotes the timestep, \( x_t \) represents the input, \( h_t \) signifies the hidden state that stores all historical information, and \( y_t \) indicates the output. For simplicity, we denote the Mamba module by \( \mathcal{F}_{mamba} \), which takes a sequence \( x \in \mathbb{R}^{B \times L \times C} \) as input and produces the corresponding output.

\subsubsection{PSM:  Position Shifting Mamba for Coordinates Priors:} Given the relatively fixed anatomical position of the CC, we employ Mamba modules to accurately predict the true coordinates from the manually initiated coordinates and corresponding high-dimensional features. This approach leverages the inherent spatial consistency of the CC to refine the initial estimations. For simplicity, we denote the Mamba module by \( \mathcal{F}_{mamba} \), which takes a sequence \( x \in \mathbb{R}^{B \times L \times C} \) as input and produces the corresponding output. Formally, let $\mathbf{p}_i^{\text{init}}$ represent the initial coordinates of the $i$-th CC's landmarks, with $\mathbf{p} \in \mathbb{R}^{B \times N_v \times 3}$, where $B$ is the batch size, $N_v$ is the number of landmarks, and $3$ represents the x, y and z coordinates. 

\begin{table*}[]
\centering
\caption{Comparison of various segmentation methods on the CCSeg Test and OOD Testset. Methods referenced: ResUNet \citep{ronneberger2015u}, nnUNet \citep{isensee2021nnu}, TwoStage \citep{gong2024intensity}, LViT \citep{li2023lvit}, SwinUnetr \citep{hatamizadeh2021swin}, Unetr++ \citep{shaker2024unetr++}, nnMamba \citep{gong2024nnmamba}. The \textit{p-value} is calculated between the best performed method and other methods. `N.S.' indicates non-significant results.}
\resizebox{\textwidth}{!}{%
\begin{tabular}{@{}cccccccccccc@{}}
\toprule
\multirow{2}{*}{Method} & \multirow{2}{*}{Backbone} & \multicolumn{4}{c}{CCSeg Test}                   & \multicolumn{4}{c}{OOD Testset}                          & \multirow{2}{*}{Param (M)} & \multirow{2}{*}{Flops (G)} \\
                        &                           & DSC          & \textit{p-value} & NSD           & \textit{p-value} & DSC          & \textit{p-value}          & NSD          & \textit{p-value}          &                            &                            \\ \midrule
ResUNet                 & CNN                       & 74.4$\pm$2.8 & \textless0.001& 61.1$\pm$3.7  & \textless0.001& 56.9$\pm$2.8 & N.S.& 42.2$\pm$2.5 & 0.003            & 15.2                 & 157.0\\
nnUNet                  & CNN                       & 66.4$\pm$5.3& \textless0.001& 45.8$\pm$3.8& \textless0.001& 50.4$\pm$3.4& \textless0.001& 38.5$\pm$3.6& \textless0.001& 30.8& 525.7\\
TwoStage                & CNN                       & 69.4$\pm$3.7 & \textless0.001& 55.0$\pm$3.6  & \textless0.001& 52.9$\pm$3.0 & 0.001            & 38.8$\pm$3.6 & \textless0.001 & 12.4                 & 123.8\\
LViT                    & Transformer               & 75.1$\pm$1.3 & \textless0.001& 64.8$\pm$0.6  & \textless0.001& 56.4$\pm$3.2 & N.S.& 43.3$\pm$3.2 & 0.01             & 5.1                 & 85.6                       \\
SwinUnetr               & Transformer               & 67.2$\pm$1.2 & \textless0.001& 53.7$\pm$1.6  & \textless0.001& 50.2$\pm$2.6 & \textless0.001 & 37.5$\pm$1.7 & \textless0.001& 15.0& 184.2\\
Unetr++                 & Transformer               & 66.5$\pm$1.3 & \textless0.001& 46.6$\pm$2.5  & \textless0.001& 48.0$\pm$1.0 & \textless0.001 & 28.9$\pm$1.6 & \textless0.001& 51.0& 92.2\\
nnMamba                 & Mamba                     & 76.0$\pm$1.5 & \textless0.001& 66.3$\pm$2.9  & \textless0.001& 57.2$\pm$2.2 & N.S.& 43.1$\pm$2.7 & 0.008            & 16.0& 159.0\\
Ours                    & Mamba                     & 78.5$\pm$1.0 & -       & 74.7$\pm$1.4  & -       & 58.7$\pm$1.2 & -                & 47.8$\pm$1.2 & -                & 16.3& 159.9\\ \bottomrule
\end{tabular}%
}
\label{tab:sota}
\end{table*}

For generating initial coordinates $\mathbf{p}_i^{\text{init}}$ and ground truth coordinates $\mathbf{p}_i^{\text{gt}}$,  we take the ground truth CC segments of an image, using morphology method to get the center line of each CC, and sample evenly across the center line from one end to another to get the $N_v$ landmarks. Considering the variable lengths of the costal cartilages, we propose a variable-length coordinate regression method to assign different numbers of regression points for each CC. Specifically, we set the upper and lower limits of the number of points to M1 and M0, respectively, corresponding to the longest and shortest costal cartilages after skeletonization. The number of points for costal cartilages with lengths in between is assigned proportionally based on their lengths. First, we calculate the actual length \( L_i \) of the center line for each CC. Then, we map the length \( L_i \) of each CC to the predefined point range from M0 to M1:

\begin{equation}
    N_v = \text{M0} + \left( \frac{L_i - L_{\min}}{L_{\max} - L_{\min}} \right) \times (\text{M1} - \text{M0}),
\end{equation}
where \( L_{\min} \) and \( L_{\max} \) are the lengths of the shortest and longest costal cartilages, respectively. Finally, we evenly sample \( N_v \) points along the center line of each CC to obtain the initial coordinates \(\mathbf{p}_i^{\text{init}}\).

To predict the three-dimensional coordinates $\mathbf{p}_i^{\text{pred}}$ for the current image, let $[, ]$ be the concatnation operation, we first concatenate $\mathbf{p}_i^{\text{init}}$ to their corresponding high-dimensional features $x_c \in \mathbb{R}^{B \times N_v \times C}$ and get a sequence of  $x_f\in \mathbb{R}^{B \times N_v \times (C+3)}$, through a fully connected layer $\mathcal{F}_{\text{fc}}$ to decrease its dimensionality to $C$ :
\begin{equation}
\mathbf{f}_i =\mathcal{F}_{\text{fc}}([\mathbf{p}_i^{\text{init}} , x_c]).
\end{equation}
Next, we use the Mamba module \(\mathcal{F}_{mamba}\) to perform sequence modeling on \(\mathbf{f}_i\), followed by another fully connected layer $\mathcal{F}_{\text{fc\_out}}$ to reduce the dimensionality to three-dimensional coordinates \((x, y, z)\):
\begin{equation}
\mathbf{p}_i^{\text{pred}} = \mathcal{F}_{\text{fc\_out}}(\mathcal{F}_{mamba}(\mathbf{f}_i)).
\end{equation}
By directly predicting the coordinates $\mathbf{p}_i^{\text{pred}}$, we can estimate the true anatomical positions of the CC. The ground truth coordinates $\mathbf{p}_i^{\text{gt}}$ are used to supervise the training of the Mamba module. The features extracted from these predicted coordinates $\mathbf{p}_i^{\text{pred}}$ are then utilized in subsequent interaction steps, enabling the model to capture more precise and contextually relevant information. This process is crucial for enhancing the overall accuracy and robustness of the segmentation framework.

For the Topology Guided Deformable Mamba Module, we employ a bipartite graph matching loss~\citep{carion2020end} to align the predicted and ground truth centerline points. The loss is defined as follows:
\begin{equation}
\mathcal{L}_{\text{match}} = \sum_{v=1}^{20} \min_{\sigma \in S_N} \sum_{i=1}^{N_v} \left\| \mathbf{p}_i^{\text{pred}} - \mathbf{p}_{\sigma(i)}^{\text{gt}} \right\|,
\end{equation}
where \( S_N \) denotes the set of all permutations of \( N_v \) elements, and \( \sigma \) represents a specific permutation. $\mathcal{L}_{\text{match}}$ ensures that each predicted point is matched optimally to a ground truth point, minimizing the overall alignment cost.

\subsubsection{GDM: Grouped Deformable Mamba for Long-range Relationship Modeling} To effectively model the long-range interactions of the CC structure, we develop the Grouped Deformable Mamba (GDM) that contains 20 distinct Mamba modules, each dedicated to the specific CC segment.  By doing so, our TGDM network can better understand the CC's morphology and topology. This method is detailed in Figure~\ref{fig:attn}.

Based on the predicted coordinates $\mathbf{p}_i^{\text{pred}}$ from the previous PSM, we extract features from the feature map $\mathbf{x}_{\text{feature}}$ and use the GDM to model the long-range relationships. Firstly, we extract the features from the feature map $\mathbf{x}_{\text{feature}}$ with predicted coordinates $\mathbf{p}_i^{\text{pred}}$. After that, we send the features to the GDM. This process can be formulated as follows: Let $[\cdot]$ be the indexing operation, $\mathcal{F}_{\text{fc\_in}}$ be the fully connected layer, and $\mathcal{F}_{mamba}$ be the Mamba module. The resulting feature sequence has the shape $(20, C, N)$, where $20$ is the number of CC segments, $C$ is the number of feature channels, and $N$ is the number of points on the centerline of the CC segments, defined in $\mathbf{p}_i^{\text{pred}}$. This $N$ can be modified to adapt to our model:
\begin{equation}
\mathbf{f}_i^{\text{mamba}} = \mathcal{F}_{mamba}(\mathcal{F}_{\text{fc\_in}}(\mathbf{x}_{\text{feature}}[\mathbf{p}_i^{\text{pred}}])),
\end{equation}
where $\mathbf{f}_i^{\text{mamba}} \in \mathbb{R}^{20 \times C \times N}$. Next, integrate the Mamba module's output back into the original feature map using a residual connection. Let $20$ be the number of CC segments., the aggregated feature set $\mathbf{x}_{\text{feature}}^{\text{res}}$ is computed as:
\begin{equation}
\mathbf{x}_{\text{feature}}^{\text{res}} = \mathbf{x}_{\text{feature}} + \sum_{i=1}^{20} \mathbf{f}_i^{\text{mamba}}.
\end{equation}

This residual connection ensures that the model retains the original feature information while incorporating the long-range dependencies captured by the Mamba modules. The aggregated feature set $\mathbf{x}_{\text{feature}}^{\text{res}}$ is then reintroduced into the backbone (lower part in Figure~\ref{fig:pipeline}), forming an iterative feedback loop that allows the model to continuously refine its understanding and update its predictions based on the long-range relationships modeled by the Mamba modules. This iterative feedback mechanism ensures that the segmentation framework can accurately delineate the CC, even in regions with low contrast or ambiguous boundaries, thereby significantly enhancing the overall segmentation DSC and NSD scores.

\subsubsection{Overall loss function and network structure}
This coarse segmentation $\mathbf{S}_{\text{low}}$ provides a reliable estimate of the cartilage location, which is then used to initialize the coordinates for the subsequent high-resolution segmentation. Let $\mathbf{X}_{\text{cropped}}$ be the high-resolution input image cropped and resized based on the predicted bounding box from the first stage, and $\mathbf{Y}_{\text{cropped}}$ be the corresponding ground truth segmentation adjusted to fit within this predicted bounding box. The final high-resolution segmentation process can be described as:
\begin{equation}
\mathbf{S}_{\text{cropped}} = \mathcal{F}_{\text{tgdm-unet}}(\mathbf{X}_{\text{cropped}}, \mathbf{S}_{\text{low}}).
\end{equation}
Here, $\mathbf{S}_{\text{low}}$ serves as an initial guide for generating $\mathbf{X}_{\text{cropped}}$ and $\mathbf{Y}_{\text{cropped}}$, thereby improving the accuracy and precision of the final segmentation result. In the second stage, the loss function for optimizing the $\mathcal{F}_{\text{tgdm-unet}}$ network, using the cropped high-resolution input image $\mathbf{X}_{\text{cropped}}$, the corresponding ground truth segmentation $\mathbf{Y}_{\text{cropped}}$, and the predicted segmentation $\mathbf{S}_{\text{cropped}}$, is given by:
\begin{equation}
\mathcal{L}_{\text{seg}} = \mathcal{L}_{\text{Dice}}(\mathbf{S}_{\text{cropped}}, \mathbf{Y}_{\text{cropped}}) + \mathcal{L}_{\text{CE}}(\mathbf{S}_{\text{cropped}}, \mathbf{Y}_{\text{cropped}}).
\end{equation}

The overall loss is the sum of the segmentation and alignment losses:
  \begin{equation}
  \mathcal{L}_{\text{stage2}} = \mathcal{L}_{\text{seg}} + \mathcal{L}_{\text{match}}.
  \end{equation}


\begin{table*}[!ht]
\centering
\caption{Demographic and Clinical Characteristics of ours CCSeg benchmark. This table summarizes the distribution of training, validation, and test samples across different age groups in the CCSeg. Each group is defined by specific clinical characteristics related to the calcification of the CC, ranging from minimal or absent calcification in children and adolescents to mild or extensive calcification in adults and individuals with cartilage harvested.}
\label{tab:stat}
\begin{tabular}{lp{9.5cm}ccc}
\toprule
\textbf{Groups}                & \textbf{Characteristic}                                                                                                                        & \textbf{Training}   & \textbf{Validation} & \textbf{Test}       \\ \midrule
Group 1                        & \multirow{2}{=}{\raggedright school-aged children, with minimal or absent calcification of the CC}                               & \multirow{2}{*}{21} & \multirow{2}{*}{10} & \multirow{2}{*}{10} \\
Ages 6-11                      &                                                                                                                                               &                     &                     &                     \\ \midrule
Group 2                        & \multirow{2}{=}{\raggedright adolescents, with minimal or absent calcification of the CC}                                        & \multirow{2}{*}{23} & \multirow{2}{*}{10} & \multirow{2}{*}{10} \\
Ages 12-17                     &                                                                                                                                               &                     &                     &                     \\ \midrule
Group 3                        & \multirow{2}{=}{\raggedright adults, with mild to moderate calcification of the CC}                                              & \multirow{2}{*}{21} & \multirow{2}{*}{10} & \multirow{2}{*}{10} \\
Ages 18 and above              &                                                                                                                                               &                     &                     &                     \\ \midrule
Group 4                        & \multirow{2}{=}{\raggedright individuals exhibiting a partial absence of CC and potential extensive calcification in the surgical area} & \multirow{2}{*}{20} & \multirow{2}{*}{10} & \multirow{2}{*}{10} \\
Cartilage harvested &                                                                                                                                               &                     &                     &                     \\ \midrule
Total                          & -                                                                                                                                             & 85                  & 40                  & 40                  \\ \bottomrule
\end{tabular}
\end{table*}

\section{CCSeg: Benchmarking Costal Cartilage Segmentation}

\subsection{Motivation}
CC segmentation is challenging because the foreground intensity is similar to the background intensity (such as liver and intercostal muscles) \citep{weber2021}, especially for younger teenagers since their CC is softer than that of adults \citep{Wang2024}. Currently, there are no publicly available datasets or benchmarks for CC segmentation.
Thus, to advance the research on the computer-aided diagnosis and surgery system for the CC, we contribute  \textbf{CCSeg}, a benchmark for CC segmentation from CT scans. CCseg contains 165 cases with accurate voxel-vise annotation of each costs cartilage.  The details of our benchmark are shown in Table 1.
Ethical approval of the study was obtained from the institutional review board of the Plastic Surgery Hospital of Chinese Academic of Medical Science and Peking Union Medical College, Beijing, China [Serial number 2023(219)].

\subsection{Data Correction and Annotation}

We have established the following inclusion and exclusion criteria for our study on CC segmentation following \citet{han2015, zhou2016, velzen2020}. The details of there criterias and annotation steps are shown below:

\begin{enumerate}
    \item \textbf{Patient Selection:}
        (1) Primary cohort: Patients (6 years and older) with congenital or acquired ear deformities who were scheduled for ear reconstruction surgery at the Plastic Surgery Hospital, Chinese Academy of Medical Sciences, and who had undergone chest CT scans.
        (2) External test cohort: Patients who underwent chest CT scans at the Second Hospital of Hebei Medical University.
    \item \textbf{CT Scan Requirements:}
        (1) Plastic Surgery Hospital: CT scans should ideally capture a complete scan of the 1st to 10th CCs. 
        (2) Second Hospital of Hebei Medical University: There are no specific requirements for the completeness of CT scans.
    \item \textbf{Exclusion Criteria:}
        (1) Patients are excluded if the imaging quality is poor or significantly affected by severe artifacts.
        (2) Patients with severely calcified CC are excluded.
        (3) Patients with severe psychological or physical illnesses.
    \item \textbf{Annotation Steps:}
    Four plastic surgery residents independently performed manual segmentation of each patient's entire set of scan images under the supervision of radiology experts. The steps involved were as follows: (1) DICOM files were loaded into 3D Slicer, and the 'segment editor' module was opened. (2) New segments were created and named, e.g., “costal cartilage L1”. (3) Segmentation was performed using interpolation from the sagittal or coronal plane. (4) Manual tools such as 'paint' and 'erase' were used for refinement. (5) The segmentation was reviewed in 3D view and across axial, coronal, and sagittal planes and adjusted as necessary. (6) Finally, all scan image labels were corrected by a senior plastic surgeon under the guidance of a radiology expert to ensure accuracy and consistency.
\end{enumerate}

In this study, high-resolution chest CT imaging was performed at the Plastic Surgery Hospital of the Chinese Academy of Medical Sciences and Peking Union Medical College using two Philips-manufactured scanners: the Philips CT 6000 and the Philips Brilliance 64. The CT data spans from September 2014 to October 2023. Both scanners were configured with a slice thickness of 1 mm and an image resolution of 512×512 mm\textsuperscript{2}. The Philips CT 6000 operated at kilovoltage peaks (KVP) of 100, 120, or 140 kV, with a tube current ranging from approximately 200 to 600 mA and a slice interval of 0.5 mm, with an image size of 523 KB. The Philips Brilliance 64 was set to a KVP of 120 kV, with a tube current range of approximately 100 to 500 mA and a slice interval of -0.5 mm or -0.6 mm, with an image size of 527 KB. We categorized participants based on developmental characteristics and whether CC had been surgically removed. Subsequently the dataset was divided into training, validation, and test sets with 85, 40, and 40 patients, respectively.

Additionally, we prepared an out-of-domain test dataset (OOD Testset) consisting of 10 patients from the Second Hospital of Hebei Medical University. The CT data for this group spans from April 2021 to January 2024, with participants aged between 8 and 27 years using GE-manufactured CT equipment, with a slice thickness and slice interval of 1.25 mm, a KVP of either 100 or 120 kV, and a tube current ranging from 71 to 307 mA, with an image resolution of 508×508 mm² and an image size of 549 KB . 

\begin{figure*}[!thbp]
\includegraphics[width=1\textwidth]{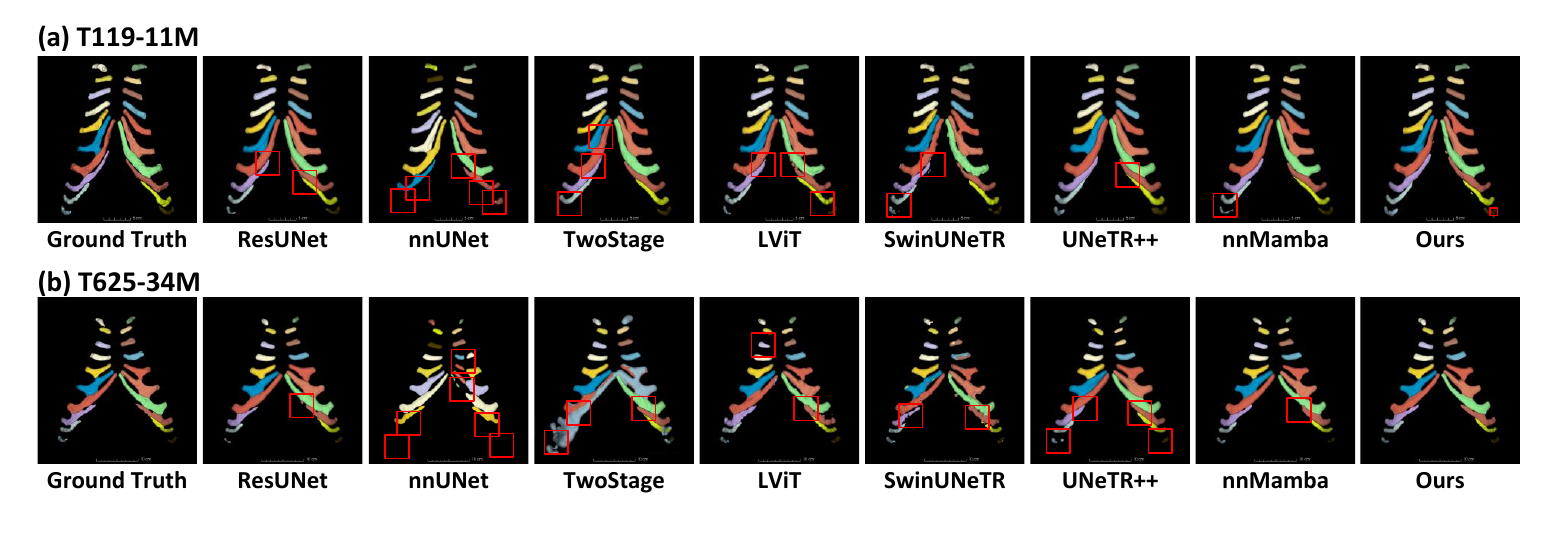}
\caption{Visual comparisons between our method and other state-of-the-art methods on CCSeg test set. Different categories of CC are denoted by different colors. The wrong prediction are marked by red boxes.}
\label{fig:quali}
\end{figure*}

\section{Experiments}
\subsection{Implementation and Evaluation Metrics}
Our models are developed using the PyTorch 2.2.0 framework, which is based on CUDA 12.1, and are trained on NVIDIA RTX 4090D GPUs equipped with 24GB of memory. In the initial phase, we design a lightweight U-Net to crop the Region of Interest (ROI). This U-Net consists of three encoder layers and three decoder layers, with the first layer containing 16 filters (i.e., channels). The lightweight U-Net is supervised by the bounding box of ground truth (GT) segments with dimensions \(64 \times 64 \times 64\). In the subsequent phase, we use our TGDM network, which consists of four encoder-decoder layers. The first layer of this network utilizes 32 filters (channels) to process inputs with dimensions \(128 \times 128 \times 128\). The backbone ResUNet architecture is described in \citep{gong2024intensity}. All models are trained for 500 epochs using the Adam optimizer, with a learning rate of \(1 \times 10^{-4}\) and a batch size of 4. We performs the three fold cross-validation on the CCSeg training-validation set to obtain the results.

We utilize two key metrics to evaluate our segmentation algorithms, as recommended by recent studies \citep{Reinke2023UnderstandingMP}: the average Dice Similarity Coefficient (DSC) and Normalized Surface Dice (NSD). The DSC measures the overlap between the predicted and GT segmentation masks. The NSD evaluates the conformity of the surface contours of the predicted segmentation.

\subsection{Comparison with State-of-the-art methods}

In this study, we compare our proposed method against several state-of-the-art segmentation approaches, including both CNN- and transformer-based models, as shown in Table~\ref{tab:sota}. As shown in the table, our method achieves the highest DSC and NSD scores of 78.5 and 74.7, respectively. This indicates a significant improvement over traditional CNN architectures like ResUNet and nnUNet, as well as Transformer-based models like LViT and SwinUnetr.

Furthermore, our method achieves this superior DSC and NSD scores while maintaining a relatively low number of parameters and FLOPs. Specifically, our model has only 16.3 million parameters and requires 159.9 gigaflops, which is more efficient compared to nnUNet's 30.8 million parameters and 525.7 gigaflops. This efficiency, combined with high accuracy, makes our approach highly suitable for practical applications where computational resources are limited. Overall, the results clearly demonstrate the effectiveness and efficiency of our proposed method in medical image segmentation tasks. We also provide the visualization results in Table~\ref{fig:quali}.

\subsection{Ablation Study}
\begin{table}[]
\centering
\caption{Ablation study measured by the DSC and NSD. Stage 1 includes direct binarization (DB) and bounding box binarization (BB), while Stage 2 includes TGDM (with a fixed number of points for each CC) and variable points number (VPN). \textit{p-value} is calculated between M4 and other methods. `N.S.' indicates non-significant results.}
\resizebox{\columnwidth}{!}{%
\begin{tabular}{@{}ccccccccc@{}}
\toprule
\multirow{2}{*}{Method} & \multicolumn{2}{c}{Stage 1}    & \multicolumn{2}{c}{Stage 2} & \multirow{2}{*}{DSC} & \multirow{2}{*}{\textit{p-value}} & \multirow{2}{*}{NSD} & \multirow{2}{*}{\textit{p-value}} \\
                        & DB& BB& TGDM         & VPN&                      &                                   &                      &                                   \\ \midrule
M0                      &              &              &              &              & 74.4$\pm$2.8         & \textless 0.001                              & 61.1$\pm$3.7         & \textless 0.001                              \\
M1                      & $\checkmark$ &              &              &              & 76.0$\pm$1.9         & \textless 0.001                              & 71.3$\pm$1.8         & \textless 0.001                              \\
M2                      &              & $\checkmark$ &              &              & 76.7$\pm$1.8         & 0.001                             & 73.7$\pm$2.0         & N.S.\\
M3                      &              & $\checkmark$ & $\checkmark$ &              & 78.3$\pm$0.7         & N.S.                             & 74.2$\pm$2.0         & N.S.                             \\
M4                      &              & $\checkmark$ & $\checkmark$ & $\checkmark$ & 78.5$\pm$1.0         & -                                 & 74.7$\pm$1.4         & -                                 \\ \hline
\end{tabular}%
}
\label{tab:abla}
\end{table}
The ablation study results, summarized in Table \ref{tab:abla}, evaluate the impact of different components on segmentation DSC scores. The analysis is divided into two stages: Stage 1 includes direct binarization (DB) and bounding box binarization (BB), while Stage 2 includes TGDM (with a fixed number of points for each CC) and variable points number (VPN). The results show that M0, which serves as a baseline without any of the components, achieves a DSC of 74.4. Incorporating DB (M1) results in a DSC of 76.0, while adding BB (M2) increases the DSC to 76.7. Adding TGDM to BB (M3) further improves the DSC to 78.3. The best result is achieved by M4, which includes both BB and TGDM with VPN, resulting in a DSC of 78.5. The \textit{p-value} comparisons indicate statistical significance between M4 and other methods, except for M3 in terms of DSC. For NSD, M4 also achieves the highest score of 74.7, with significant differences observed in M0 and M1.



\section{Discussion and Conclusion}
Our proposed Topology Guided Deformable Mamba method for CC segmentation demonstrates significant advancements over existing techniques by incorporating topological priors into a deformable model framework. This integration allows long-range relationships to be effectively captured within the complex anatomy of CC, which is essential for accurate diagnosis and surgical planning. The deformable nature of our model enhances its adaptability to diverse anatomical variations, improving its robustness across different patient datasets. Extensive experimental validation on both in-domain and out-of-domain datasets demonstrated that our approach significantly improves segmentation precision and robustness, underscoring its potential to set new standards for CC segmentation.

In conclusion, the Topology Guided Deformable Mamba method offers a novel and effective approach to the challenging task of CC segmentation. By leveraging topological priors and a deformable model, our method achieves superior adaptability and accuracy, as validated by a comprehensive benchmark of numerous cases. The significant improvements in segmentation precision and robustness over existing methods highlight the efficacy of our approach. This benchmark provides a valuable resource for future research and sets a new standard for evaluating CC segmentation techniques. Future work will extend this method to other anatomical structures and further enhance its DSC and NSD scores and generalizability, contributing to improved clinical outcomes in medical imaging applications.

\section*{Acknowledgments}
This work is supported by the National Major Disease Multidisciplinary Diagnosis and Treatment Cooperation Project [21025] and Beijing Municipal Science \& Technology Commission [Z221100007422084].

\bibliographystyle{model2-names.bst}\biboptions{authoryear}
\bibliography{refs}

\end{document}